\documentclass[10pt, conference, final, letterpaper]{IEEEtran}

\IEEEoverridecommandlockouts

\usepackage{cite}
\usepackage{amsmath}
\usepackage{amssymb}
\usepackage{amsfonts}
\usepackage{algorithm}
\usepackage{algpseudocode}
\usepackage{textcomp}
\usepackage[utf8]{inputenc}
\usepackage[english]{babel}
\usepackage[pdftex]{graphicx}
\usepackage{pgfplots}
\pgfplotsset{compat=1.17}
\usepackage{pgfplotstable}
\usepackage{filecontents}
\usepackage{subcaption}
\usepackage{comment}
\usepackage{booktabs}
\usepackage{eso-pic}

\newcommand{\lFig}[1]{\label{fig:#1}}
\newcommand{\lAlg}[1]{\label{alg:#1}}
\newcommand{\rFig}[1]{Fig. \ref{fig:#1}}
\newcommand{\lSec}[1]{\label{sec:#1}}
\newcommand{\rSec}[1]{Section \ref{sec:#1}}
\newcommand{\lTable}[1]{\label{tab:#1}}
\newcommand{\rTable}[1]{Table \ref{tab:#1}}
\newcommand{\rAlg}[1]{Algorithm \ref{alg:#1}}

\usepackage{xcolor}
\def\BibTeX{{\rm B\kern-.05em{\sc i\kern-.025em b}\kern-.08em
    T\kern-.1667em\lower.7ex\hbox{E}\kern-.125emX}}
\begin{document}




\title{Edge-based fever screening system over private 5G
{\footnotesize \textsuperscript{}}
}

\author{\IEEEauthorblockN{Murugan Sankaradas, Kunal Rao, Ravi Rajendran, Amit Redkar and Srimat Chakradhar}
\IEEEauthorblockA{
\textit{NEC Laboratories America, Inc.}\\
Princeton, NJ}
}

\newcommand\AtPageUpperMycenter[1]{\AtPageUpperLeft{%
 \put(\LenToUnit{0.05\paperwidth},\LenToUnit{-1cm}){%
     \parbox{0.8\textwidth}{\raggedleft\fontsize{9}{11}\selectfont #1}}%
 }}%
\newcommand{\conf}[1]{%
\AddToShipoutPictureBG*{%
\AtPageUpperMycenter{#1}
}
}

\maketitle
\conf{The Sixth ACM/IEEE Symposium on Edge Computing (SEC 2021)}

\begin{abstract}
Edge computing and 5G have made it possible to perform analytics closer to the source of data and achieve super-low latency response times, which isn't possible with centralized cloud deployment. In this paper, we present a novel fever screening system, which uses edge machine learning techniques and leverages private 5G to accurately identify and screen individuals with fever in real-time. Particularly, we present deep-learning based novel techniques for fusion and alignment of cross-spectral visual and thermal data streams at the edge. Our novel Cross-Spectral Generative Adversarial Network (CS-GAN) synthesizes visual images that have the key, representative object level features required to uniquely associate objects across visual and thermal spectrum. Two key features of CS-GAN are a novel, feature-preserving loss function that results in high-quality pairing of corresponding cross-spectral objects, and dual bottleneck residual layers with skip connections (a new, network enhancement) to not only accelerate real-time inference, but to also speed up convergence during model training at the edge. 
To the best of our knowledge, {\it this is the first technique} that leverages 5G networks and limited edge resources to enable real-time feature-level association of objects in visual and thermal streams (30 ms per full HD frame on an Intel Core i7-8650 4-core, 1.9GHz mobile processor). To the best of our knowledge, this is also {\it the first system} to achieve real-time operation, which has enabled fever screening of employees and guests in arenas, theme parks, airports and other critical facilities. By leveraging edge computing and 5G, our fever screening system is able to achieve 98.5\% accuracy and is able to process $\sim$ 5X more people when compared to a centralized cloud deployment.
\end{abstract}


\section{Introduction}
\lSec{introduction}
Fever is a very common symptom for various infectious diseases like COVID-19. Identifying and isolating individuals with fever, before they come in contact with others, helps in reducing the transmission and spread of the virus. Accurate fever detection requires measuring temperature at the inner canthus of the eye. To do this without human intervention, sensors with visual and thermal camera are deployed. Visual camera feed is used to identify inner canthus region while thermal camera feed is used to obtain the temperature at this region.



\rFig{edge-system-overview} shows an overview of fever screening system to detect elevated body temperature of persons entering facilities like factories, office complexes, schools, etc. Visual and thermal feed is transmitted over private 5G network, which has high bandwidth and low latency, and all analytics is performed on the edge computing infrastructure, which avoids the delays incurred in moving raw data streams to the cloud over poor bandwidth and high latency network. 

One thing to note is that the visual and thermal cameras are disparate and placed right next to each other, which makes them have slightly different view point. Due to this, both the streams have to be fused in real-time to measure the temperature accurately and for this, we use edge machine learning techniques. The measured temperature also varies depending on the distance from the sensors. Farther the distance, lesser is the measured temperature because gases and particles in the atmosphere absorb some of the emitted infrared radiation. Therefore, our system detects individuals and estimates their depth and offset in real-time to compensate for this variation. Fusing multiple sensor streams is a non-trivial task. {\it First}, fields of view are not aligned due to the physical displacement of the sensors. {\it Second}, the semantic information and data formats of the sensors are radically different \cite{camera-lidar-fusion}. 
{\it Third}, cross-spectral object localization and depth perception under application-specific affordable cost and resource constraints is a major challenge. Our key contributions in this paper are:


\begin{figure}[t]
\centering
    \includegraphics[width=0.95\linewidth]{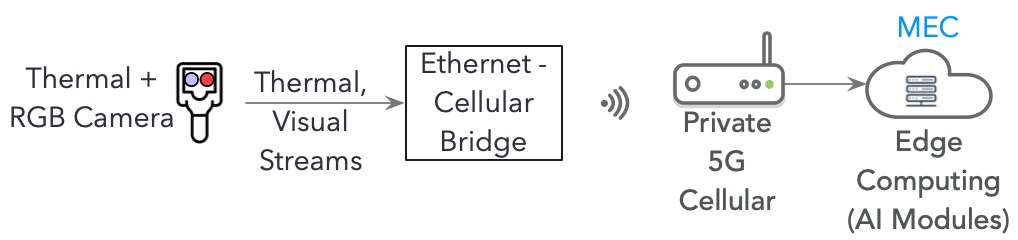}
    \caption{Fever screening system overview}
    \lFig{edge-system-overview}
\end{figure}

\begin{figure}[t]
    \centering
    \includegraphics[width=0.95\linewidth]{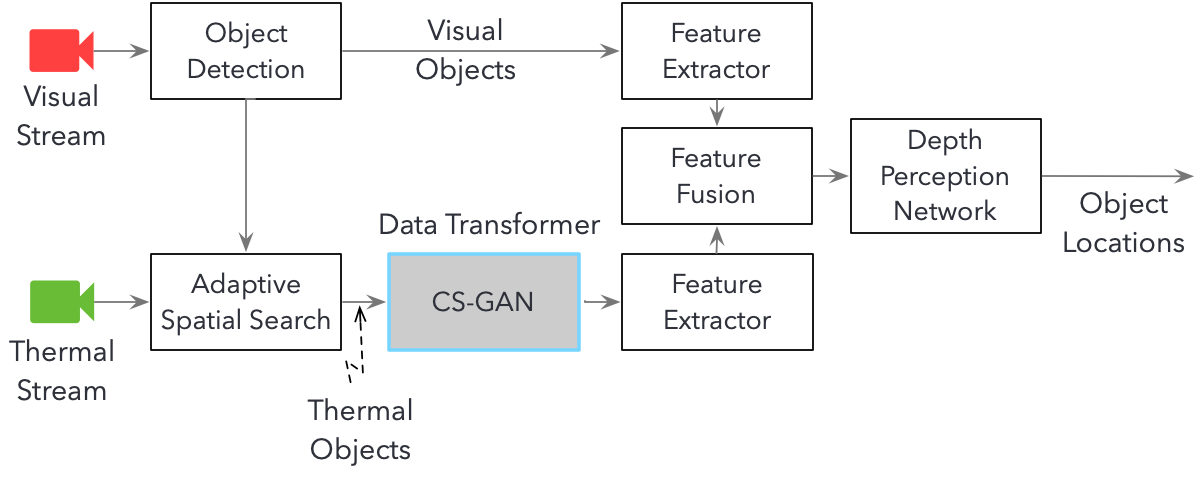}
    \vskip -0.1in
    \caption{Overview of AI Modules in Edge System}
    \lFig{intro-overview}
\end{figure}
\vskip -0.1in

\begin{itemize}
    \itemsep0em
    \item A new cross-spectral deep-learning generative adversarial network CS-GAN that synthesizes feature preserving images from thermal data, to perform real-time object feature-level association in visual and thermal streams at the edge (30 ms per full HD frame on i7-8650 CPU). 
\item A new multivariable linear regression model, to estimate location using object's feature-level correspondence. 

\item A novel edge-based fever screening system, which uses CS-GAN to associate objects in visual and thermal streams, and measures temperatures of individuals in realtime.  
\end{itemize}


Experimental results show that our fever screening system, running at the edge, is able to achieve 98.5\% accuracy and process $\sim$ 5X more people per minute compared to a centralized cloud deployment.

\section{Proposed approach}
\lSec{proposed-approach}
\setlength{\abovedisplayskip}{3pt}
\setlength{\belowdisplayskip}{3pt}
\rFig{intro-overview} shows an overview of the AI modules within the fever screening system. We propose a novel cross-spectral generative adversarial network, called CS-GAN, which is a strategically modified version of CycleGAN \cite{Zhu:cycleGAN}. Here, the input visual stream is our target domain, and we  use  known  object  detectors  available  for  this  domain  to detect  objects  like  face  or  person. We use the bounding box information of the visual object to form several candidate bounding box proposals in the thermal domain by using {\it adaptive spatial search}, which is discussed in \rSec{depth-and-offset-estimator}. This obviates the need for good object detectors in the thermal domain, where accurate object detection is not possible due to texture-less data. Since we only have good feature extractors for the target domain (visual) at our disposal, we transform images from the source domain (thermal) to the target domain (visual) by using CS-GAN.




\subsection{Cross-Spectral GAN (CS-GAN)}
CS-GAN has two networks. \rFig{thermal-to-visual-gan} shows the first network, which  synthesizes visual images from thermal data. Second network transforms visual images into the thermal domain. Our experimental results (reported later) show CS-GAN can achieve state-of-the-art generation quality, with lower Frechet Inception Distance.
\begin{figure}[H]
    \centering
    \includegraphics[width=0.95\linewidth]{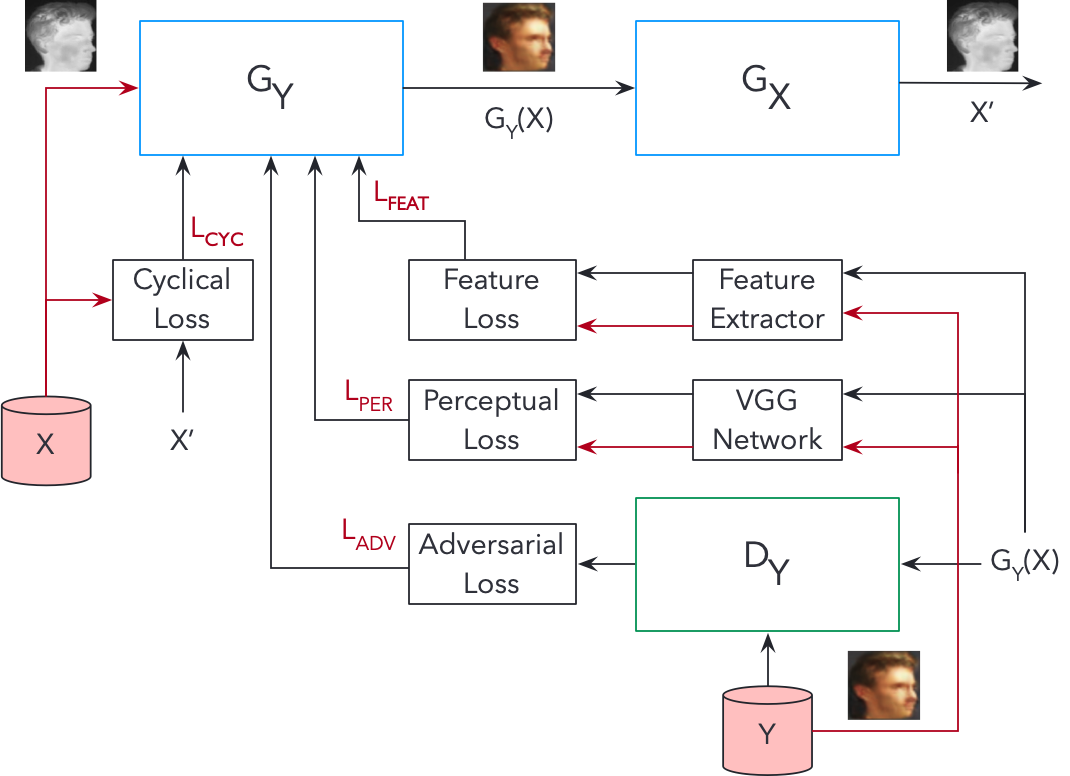}
    \vskip -0.1in
    \caption{Thermal to visual synthesis}
    \centering
    \lFig{thermal-to-visual-gan}
\end{figure}
\vskip -0.1in
\subsubsection{Thermal to visual synthesis}
\lSec{thermal-to-visual-gan}
\rFig{thermal-to-visual-gan} shows our network to synthesize visual images from thermal data. Given a thermal patch (bounding-box), our generator synthesizes visual images that satisfy a key goal: synthesized images should conserve spatial information in the thermal patch, and retain key object features and landmarks. On the other hand, our discriminator learns to judge whether the synthesized visual images are visually homogeneous and have key object features.

Given a thermal image $X$, the generator $G_Y$ synthesizes a synthetic visual image $G_Y(X)$. Synthetic visual images are used for training the discriminator $D_Y$, which is able to distinguish between the original visual image $Y$ and the synthesized visual image $G_Y(X)$. Discriminator network is able to predict whether the synthesized image is real or fake, and its output allows for computation of adversarial loss, which is required to train the discriminator and generator networks. Generator $G_X$ is used to reconstruct the original thermal image from the synthesized visual image $G_Y(X)$. Reconstructed thermal image is $X^{\prime} = G_X(G_Y(X))$. The difference between the original thermal image and the synthesized thermal image (i.e. between $X, X^{\prime}$) is used to compute the cyclical loss, which is necessary to train generator networks $G_Y\ \&\ G_X$ for cyclic consistency.

\subsubsection{Visual to thermal synthesis}
This process also uses a generator and a discriminator. Again, we use adversarial loss to train generator $G_X$ and discriminator $D_X$. For cyclical loss, we calculate the $L1$ norm between the real $Y$ and the reconstructed visual image $Y^{\prime} = G_Y(G_X(Y))$. Perceptual loss is calculated from real and synthesized thermal images ($X, G_X(Y)$).

\subsection {Loss functions}
\subsubsection{Overall Objective Function}
The overall objective function $L$ is a weighted sum of four loss functions. 
\begin{align}
    \label{overall-objective-fn} 
    \mathbb{L}(G_Y, G_X, D_Y, D_X) =&\ \nonumber\\
    \mathbb{L}_{\text{ADV}} (G_Y, G_X, D_Y, D_X, X, Y) \nonumber \\
    + \lambda_{cyc} \mathbb{L}_{\text{CYC}} (G_Y, G_X, X, Y) \nonumber \\
    + \lambda_{per} \mathbb{L}_{\text{PER}} (G_Y, G_X, X, Y) \nonumber \\
    + \lambda_{feat} \mathbb{L}_{\text{FEAT}} (G_Y, X, Y)
\end{align}

where $\lambda_{cyc} \ , \lambda_{per} \ \text{and} \ \lambda_{feat}$ are weights for cyclical, perceptual and feature preserving losses respectively.

Adversarial ($L_{ADV}$) and cyclical ($L_{CYC}$) loss functions are the same as the CycleGAN loss functions \cite{Zhu:cycleGAN}. The perceptual ($L_{PER}$) is similar to \cite{Johnson:perceptual-loss}. Feature-preserving ($L_{FEAT}$) loss functions are new, and it is described here.

\subsubsection{Feature-preserving loss}
\lSec{feature-preserving-loss}
To ensure that synthesized images retain important features, we propose a new {\it feature-preserving loss} function, which helps retain the higher-level object features (like the facial landmarks, for example). This also makes \textit{CS-GAN} to be feature-point-centric. This enables the generator to produce textured synthetic images, which makes it easy to detect various objects and associated landmarks with higher accuracy (\rTable{ablation-loss-functions}). 
Given a batch size $m$ and $k$ feature points, we define feature-preserving loss as follows:
\begin{align}
    FPL_i &= \sum\limits_{j=1}^{k}\ \lVert f_p(y_{i,j}) - f_p(G_Y(x_{i,j})) \rVert_2 & \nonumber \\
    \mathbb{L}_{\text{FEAT}}(G_Y, Y, X) &= 
        \begin{cases}
            \frac{1}{m} \sum\limits_{i=1}^{m} FPL_i&,\text{if } mRatio  < t_{feat} \\
            \quad 0&, \text{otherwise}
        \end{cases}
\end{align}
Here, $f_p(G_Y(x))$ and $f_p(y)$ are feature points from synthesized and real images, respectively; $f_p \in \mathbb{R}^2$ are coordinates of feature points in an image, $t_{feat}$ is the threshold beyond which loss is added and $mRatio$ is $\frac{\#\ of\ images\ with\ no\ features}{batch size}$.

\subsection{Dual bottleneck residual block}
Vanishing gradients are a common problem in deeper networks. 
Gradients start getting smaller and smaller as they are back-propagated to earlier layers due to chain multiplication of partial derivatives. 


We propose Dual Bottleneck Residual Block (Dual-BRB) (\rFig{dual-brb}), which consists of 4 convolutional blocks as $G_{\text{(1x1)}}$, $F_{\text{(3x3, 3x3)}}$ and $H_{\text{(1x1)}}$. The function $G$ squeezes number of channels reducing computation for function $F(.)$. Then function $H(.)$ expands channels similar to input channels. All blocks have full pre-activation. We have two skip connections in Dual-BRB. Inner skip connection is an identity mapping for function $F(.)$, while outer skip connection provides identity mapping for complete block.
\begin{figure}[H]
    \centering
    \includegraphics[width=0.95\linewidth,scale=0.55]{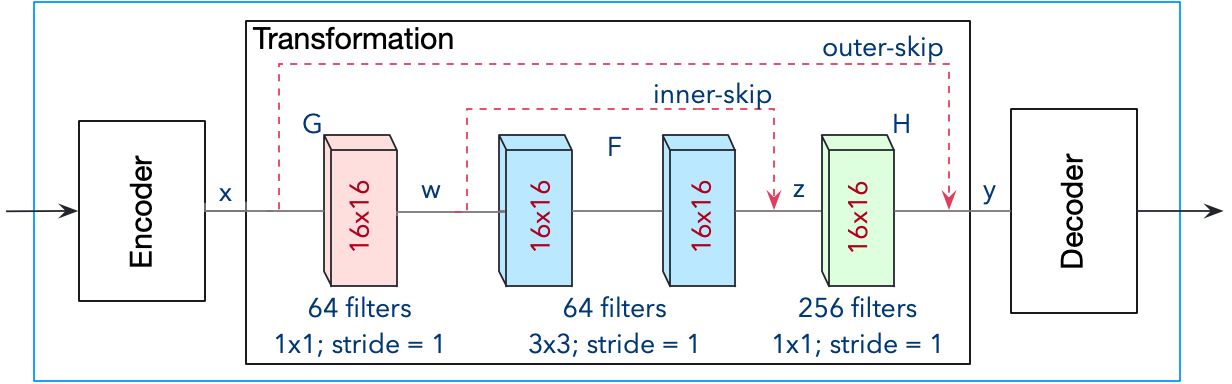}
    \vskip -0.05in
    \caption{Dual Bottleneck Residual Block}
    \centering
    \lFig{dual-brb}
\end{figure}
\vskip -0.1in
\begin{align}
    &w = G(x),\ \ z = F(w) + w \nonumber \\
    &y = H(z) + x = H(F(G(x)) + G(x)) + x \nonumber
\end{align}
where $y$ is output from Dual-BRB. Inner skip connection across $F(.)$ helps in learning residual across it, while helping in model robustness and convergence. 
\begin{figure}[H]
    \centering
    \includegraphics[width=0.95\linewidth]{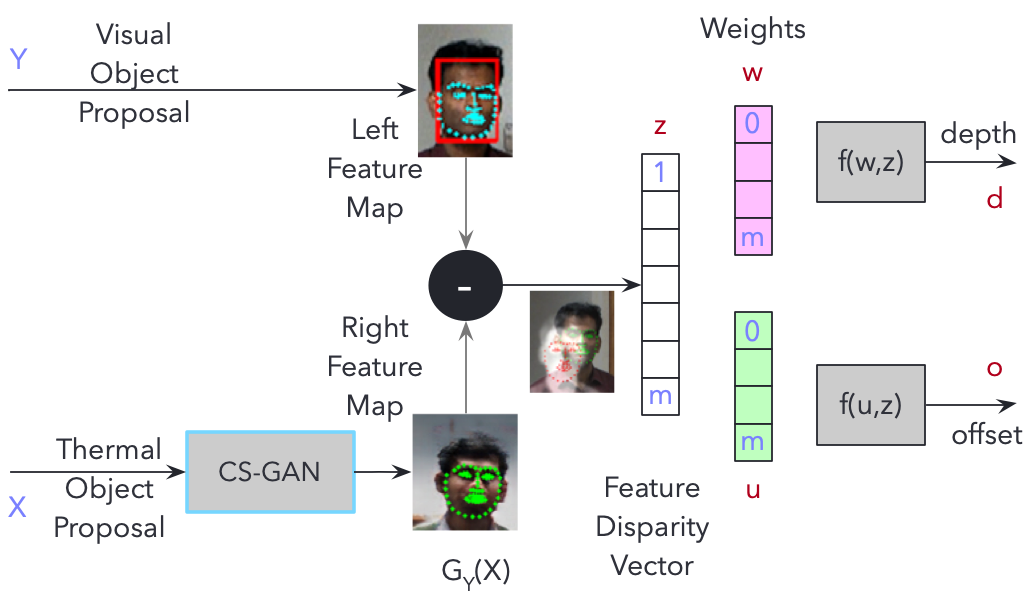}
    \vskip -0.1in
    \caption{Depth and Offset Estimator}
    \lFig{depth-and-offset-estimator}
\end{figure}
\vskip -0.1in
\subsection{Inferencing}
During inferencing, CS-GAN accepts thermal object image tiles, which are obtained from adaptive spatial search and fed to generator $G_Y$, which in turn transforms them into synthesized, visual spectrum images that retain structural information.

\subsection{Depth \& Offset Estimator}
\lSec{depth-and-offset-estimator}

\rFig{depth-and-offset-estimator} shows the procedure followed to estimate depth and offset of an individual relative to the sensor, which is required to compensate for temperature variation based on the distance. After bounding boxes of objects are identified in the visual domain, an adaptive spatial search method determines candidate object proposals in the thermal domain. This process depends on sensor displacement, sensor fields of view, zoom levels, resolutions and relative orientation.
Consider an image $Y$ with $k$ objects, $\{y_i\}_{i=1}^{k}$. 
Let the thermal image be $X$, and the thermal bounding box proposals be $B_{x_{i}} = \Phi(B_{y_{i}})$ where $\Phi$ maps a bounding box in visual domain to an estimated bounding box in thermal image. 

Adaptive search also depends on baseline (distance separating the cameras) $b$, which determines offset, angle of view and image resolution. 
We compute the function $\Phi$ by using the ratio of focal lengths of cameras and offset. Let pairs ($R_Y,\ R_X$) and ($f_Y,\ f_X$) 
represent resolution and focal length 
of visual and thermal imaging sensor. Given $\Phi\ \propto (f, R)$, a heuristic bounding box is estimated as $B_x\ =\ f_x\ B_y/f_y \pm \hat{b}$, where $\hat{b}$ is horizontal offset.
Using thermal object proposals $B_x$, visual object proposals $B_y$ are expanded, so that each visual ($y_i$) and corresponding thermal ($x_i$) cropped proposals have same size. As the next step, landmark detection is performed on $y_i$ and feature vector $\mathbf{y_i}$ is extracted. Since landmark detection cannot be performed directly on $x$, it is converted to $G_Y(x_{i})$ using CS-GAN. Landmark detection is performed on $G_Y(x_{i})$ and a feature vector $\mathbf{\hat{y}_{i}}$ is extracted.

Let $z$ be object feature disparity vector. $z$ consists of euclidean distances between $n$-feature points and angle between $n$-feature points, i.e., $z = \left( \lVert \mathbf{y} - \mathbf{\hat{y}}\rVert_{2},\ atan(\mathbf{y},\ \mathbf{\hat{y}})\right)$ where $z \in \mathbb{R}^{m}$ and $m=2n$. We regress distance ($d$) from sensors and offset ($o$) and train a multivariable linear regressor by using $2n$ explanatory variables. The regressor minimizes residual sum of squares. If the coefficients of distance-estimator model are $w \in \mathbb{R}^{m+1}$, and offset-estimator coefficients are $u \in \mathbb{R}^{m+1}$, then the object distance and offset are estimated as follows:
\begin{align}
    d = &\ w_0 + \sum\limits_{j=1}^{m} w_j z_j + \hat{\epsilon_d} = {w}^{T}\ {z} + \hat{\epsilon_d} \\
    o = &\ {u}^{T}\  {z} + \hat{\epsilon_o}
\end{align}

where $\hat{\epsilon_d}$ and ${\hat{\epsilon_o}}$ are distance and offset residuals. 


\begin{figure}[t!]
    \centering
    \includegraphics[width=0.99\linewidth]{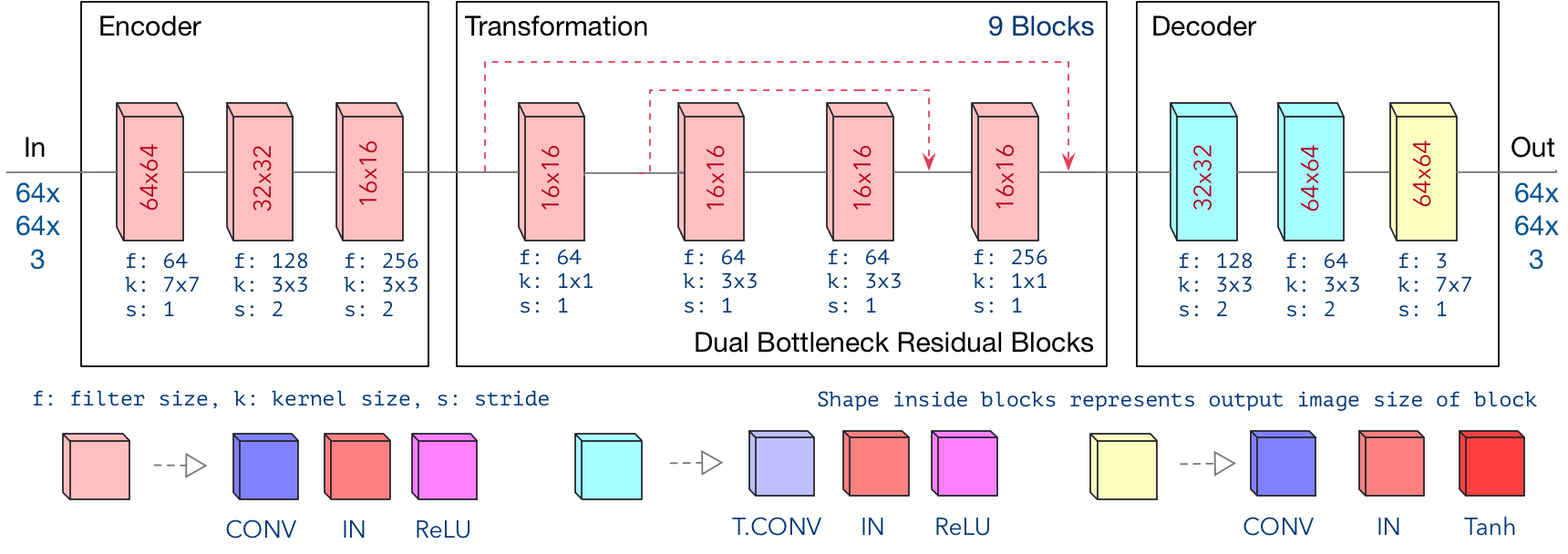}
    \caption{Network architecture of Generator }
    \centering
    \lFig{network}
\end{figure}

\begin{figure*}[t]
    \centering
    \includegraphics[width=0.7\linewidth]{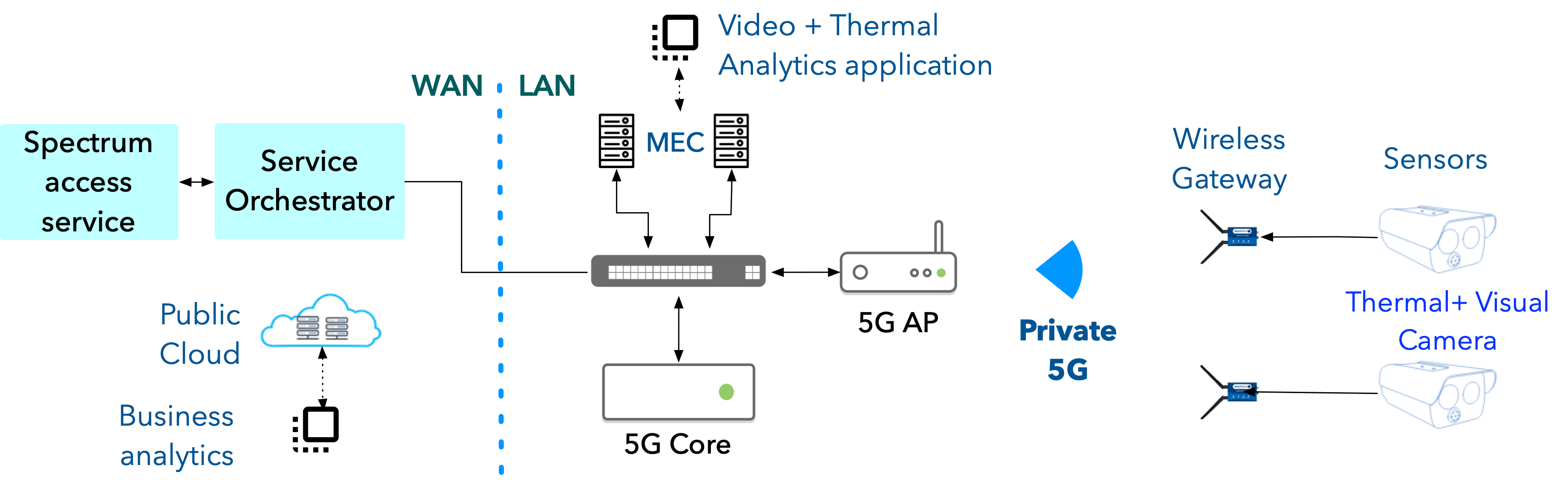}
    \caption{Deployment architecture}
    \label{deployment-architecture}
\end{figure*}

\section{Experimental setup}
\lSec{experiments}
\subsection{Compute and Communication Network setup}

Private 5G which is also known as a local or non-public 5G network, is LAN which uses 5G technologies to create a dedicated network with unified connectivity, optimized services and a secure means of communication within a specific physical area. It delivers the speed, latency and other benefits promised by 5G to applications. Our system utilizes Band 48 (CBRS) in 3.5 GHz spectrum. It spans between 3550 MHz to 3700 MHz range, under Tier 3 Priority Access License (PAL).

CBRS (Citizens Broadband Radio Service) allows economic way for businesses to deploy private LTE and 5G network in their premises. It provides better indoor and outdoor cellular signal with enhanced  security. We have used both indoor and outdoor Access Points to form local cellular network. Thermal and RGB cameras are connected to cellular network using ethernet-to-cellular bridge devices. Overview of our deployment is shown in Fig. \ref{deployment-architecture}. 

We have used combination of Multitech \cite{multitech-link} and Cradlepoint \cite{cradlepoint-link} wireless gateways to connect Customer Premise Equipment (CPE) devices over private 5G to Access Points from Celona \cite{celona-link}. These access points provide coverage for upto 25K SQFT indoor area and 1M SQFT outdoor area. Spectrum Access System (SAS) is provided by Federated Wireless \cite{federated-sas-link}. Deployment of devices at the edge is managed through Celona's Service Orchestrator, which resides in the cloud. Along with initial deployment, Service Orchestrator also provides functions for routine maintenance and any policy related automation. Our Multi-access Edge Computing (MEC) \cite{7901477} server setup consists of one master node server with 10-core Intel core i9 CPU and five worker node servers equipped with 24-core Intel CPUs.

\subsection{Generator Network architecture}
\rFig{network} shows architecture of our generator network, which consists of an encoder, a transformer and a decoder block. Encoder network consists of a 7x7 convolution layer, followed by down-sampling layers using two 3x3 convolution (stride-2). 
Dual-BRBs reduce inference time by a factor of 2 when compared with basic residual block implementation in \cite{He:resnet}, and inference speed is achieved without degrading image quality (\rFig{res-image-quality-vs-blocks}). The final decoder network consists of two up-sampling layers of 3x3 transpose convolution (T.CONV) and a final 7x7 convolution layer with $tanh$ activation. Discriminator networks $D_Y$ and $D_X$ are similar to \cite{Isola:pix2pix}.

\subsection{Training}
Training was done on an Intel Core i9-9900K CPU @ 3.6 GHz, which has an Nvidia GeForce RTX 2070 GPU with 8GB Memory. 
Image pre-processing like cropping and resizing of images was followed by normalization. An image size of 64x64 was used across all experiments. Thermal data was converted from 1 channel to 3 channels to keep it consistent with 3 channels in visual image.
We used $\lambda_{\text{cyc}}$ = 10, $\lambda_{\text{per}}$ = 4.0 and $\lambda_{\text{feat}}$ = 7.0 in the overall objective function (\ref{overall-objective-fn}). Adam Optimizer was used with parameters as $\alpha$ = 0.0002, $\beta_1$ = 0.5 and $\beta_2$ = 0.999. We start decaying learning rate based on feature preserving loss.

\subsection{Datasets}
We used proprietary data for training, and public data for validation and testing.

{\bf Proprietary data:}
We collected extensive facial images from Intel Real-sense and Mobotix M16-TR thermal camera during our field trials. Person and face image pairs are labeled with distance and offset measurements using Intel RealSense.



{\bf Public data - IRIS \cite{Abidi:iris-dataset}:} It has paired thermal images and visible images for faces with various poses, expressions and illuminations.

\begin{figure}[H]
\begin{subfigure}[]{0.5\linewidth}
    \centering
    \includegraphics[height=1.5 in]{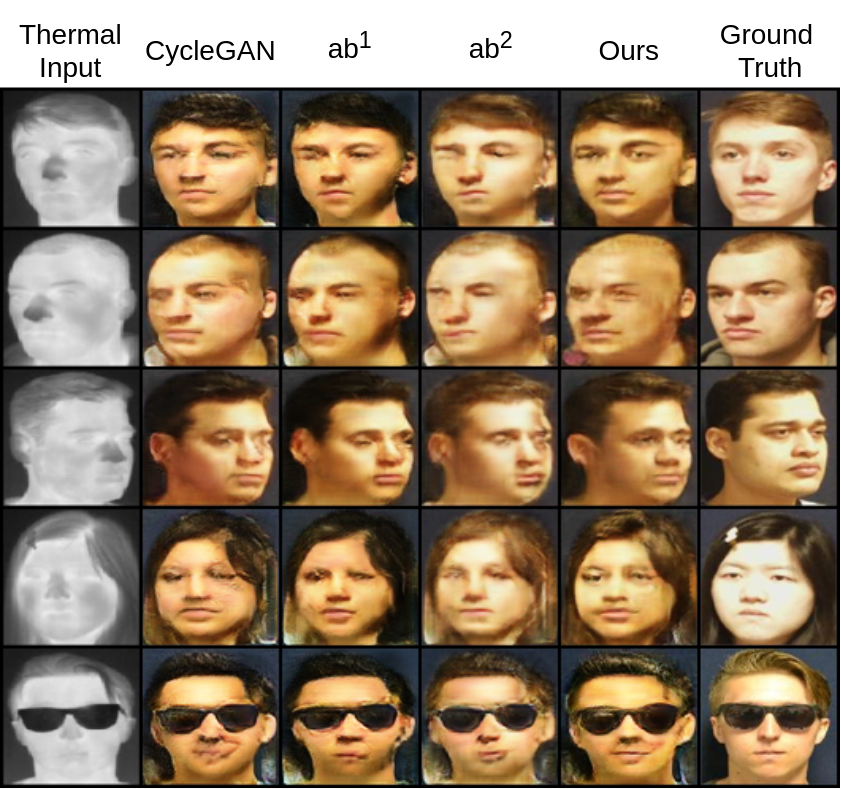}
    \caption{Image quality comparison}
    \lFig{res-image-quality-vs-blocks}
\end{subfigure}%
\begin{subfigure}[]{0.5\linewidth}
    \centering
    \includegraphics[height=1 in]{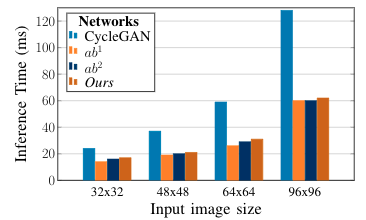}
    \caption{Inference times comparison}
    \lFig{res-inference-time-vs-transformer-blocks}
\end{subfigure}
\caption{Analysis of Dual Bottleneck Residual Blocks}
\end{figure}
\vskip -0.1in

\section{Results}
\lSec{results}
\definecolor{blue1}{HTML}{AAD8F1}
\definecolor{blue2}{HTML}{4DC6DD}
\definecolor{blue3}{HTML}{005789}

\definecolor{darkblue}{HTML}{1F77B4}
\definecolor{darkorange}{HTML}{FF7F0E}
\definecolor{darkdarkblue}{HTML}{042F66}
\definecolor{darkdarkorange}{HTML}{C26300}


\subsection{Dual Bottleneck Residual Blocks}
\textit{Qualitative analysis}: In \rFig{res-image-quality-vs-blocks}, we  compare image quality across different networks. The first network is CycleGAN with \textit{Basic Residual Block}. The second network $ab^{1}$ is an ablation of proposed network, where we use \textit{Basic Bottleneck Residual Block} instead of Basic Residual block. This usage is similar to \cite{He:resnet}. The third network $ab^{2}$ is another ablation of proposed network, but with \textit{Single Skip Bottleneck Residual Block}. The fourth network is the proposed Dual-BRBs. We could not reliably detect landmark features from all images synthesized by CycleGAN, $ab^{1}$ or $ab^{2}$ networks. However, our landmark detector was able to detect landmarks across all images synthesized by CS-GAN.

\noindent\textit{Quantitative analysis}: 
\lSec{result-quantitative-analysis}
We measured transformation times [Intel Core i7-8650U CPU @ 1.9 GHz] for various input image sizes. 
As shown in \rFig{res-inference-time-vs-transformer-blocks}, inference time is about 31 ms  to transform 64x64 thermal image into synthetic visual image. This performance is good enough for real-time applications. 

\subsection{Loss functions}
\noindent\textit{Qualitative analysis}: We evaluated following combinations of loss functions: (a) No Perception loss (P) + No Landmark Loss (LM) + No Object detector (OD) loss, (b) With P + No LM + No OD, (c) With P + No LM + With OD, (d) With P + With 5-LM + With OD, and (ours) (proposed overall loss function) With P + With 68-LM + With OD. 

\noindent\textit{Quantitative analysis}: 


\begin{table}[H]
    \caption{Impact of ablated loss functions on synthesized images}
    \lTable{ablation-loss-functions}
    \vskip -0.25in
    \begin{center}
    \scalebox{0.99}{
    \begin{tabular}{lccccc}
    Evaluation Metrics & (a) & (b) & (c) & (d) & (e) \\
    \midrule
    Face Detector Accuracy & 0.94 & 0.96 & 0.96 & 0.97 & 0.97\\
    Feature Detector Error & 28.8 & 19.7 & 19.1 & 16.7 & 16.3 \\
    FID Score & 85 & 55 & 53 & 39 & 37 \\
    \bottomrule
    \end{tabular}
    }
    \end{center}
\end{table}
\vskip -0.19in

Our results shown in \rTable{ablation-loss-functions} confirm out method is able to synthesize visual images that yield higher object detector accuracy, lower error, and lower FID score by using proposed overall loss function, when compared with other ablated versions.

\subsection{Comparison with other methods}
\lSec{comparison-of-depth-results-vs-others}
We compare our framework with three baselines: CycleGAN \cite{Zhu:cycleGAN}, TV-GAN \cite{Zhang:tv-gan}, TR-GAN \cite{Kezebou:tr-gan}. Our framework is able to generate visual object counterparts, with better object detection accuracy (0.98), with lower errors. The proposed network CS-GAN produces realistic visual images under diverse conditions in realtime, at 30ms for 64x64 image. We also compared our method of object position estimation with pixel-wise stereo disparity method using KITTI dataset, wherein total processing time is around 18.1 seconds. Using our method, it takes only 73 milliseconds per frame (using similar hardware as pixel-wise method described above), achieving real-time processing.

\pgfkeys{/pgfplots/linelabel/.style args={#1:#2:#3}{name path global=labelpath,execute at end plot={
\path [name path global = labelpositionline]
(rel axis cs:#1,0) --
(rel axis cs:#1,1);
\draw [help lines,text=black,inner sep=0pt,name intersections={of=labelpath and labelpositionline}] (intersection-1) -- +(#2) node [label={#3}] {};},
}}


\pgfplotstableread{
actual predicted-mean predicted-min  predicted-max
4 3.811 0.532 1.063 
5 4.578 0.357 0.488
6 6.319 0.033 0.727 
7 7.559 0.284 0.86
8 8.058 0.025 0.185
9 8.934 0.322 0.707
10 9.669 0.053 0.892
11 10.934 0.212 0.7
12 12.135 0.27 0.37
13 12.932 0.36 0.42
14 13.86 0.24 0.452
15 15.04 0.14 0.546
}{\actualVsPredictedDistance}

\pgfplotstableread{
predicted residuals
7.284 -0.284
7.621 -0.621
14.185 -0.15
9.626 0.374
12.185 -0.151
7.716 -0.716
13.24 -0.247
11.31 0.21
7.506 -0.506
12.598 -0.224
3.557 0.443
12.31 0.32
9.775 0.225
6.727 -0.727
3.95 0.05
9.322 -0.322
13.185 0.185
9.31 0.69
14.322 -0.322
6.54 -0.54
8.293 0.707
10.053 -0.053
9.161 -0.161
14.937 0.063
6.041 -0.041
11.108 0.12
7.364 -0.364
2.937 1.063
12.228 0.112
11.284 -0.284
13.506 -0.137
5.967 0.033
12.041 -0.041
4.532 -0.532
13.053 -0.053
11.053 -0.053
8.888 0.112
13.284 0.382
7.488 -0.488
9.473 0.527
9.974 0.026
7.975 0.025
9.714 0.286
9.983 0.017
13.293 0.21
8.185 -0.185
10.634 -0.634
12.488 -0.188
10.089 -0.089
7.86 -0.86
14.56 0.16
4.512 0.488
8.014 -0.014
4.082 -0.082
11.716 -0.716
13.557 0.243
10.31 0.32
9.108 0.892
9.187 -0.187
14.888 0.132
8.595 0.405
7.634 -0.634
10.053 -0.053
4.643 0.357
14.506 -0.106
9.089 -0.089
}{\predictedVsResiduals}

\pgfplotstableread{
actual predicted-mean predicted-min  predicted-max
0 0.3 0.3 0.6
1 1.2 0.14 0.2
2 2.14 0.532 0.3
3 3.578 0.357 0.488
4 4.01 0.033 0.727 
5 5.159 0.284 0.356
6 6.019 0.025 0.185
}{\actualVsPredictedOffsetDOne}

\pgfplotstableread{
actual predicted-mean predicted-min  predicted-max
0 0.28 0.12 0.236
1 1.3 0.14 0.2
2 2.24 0.22 0.3
3 3.78 0.357 0.288
4 3.9 0.124 0.27 
5 5.02 0.211 0.156
6 5.834 0.112 0.15
}{\actualVsPredictedOffsetDTwo}

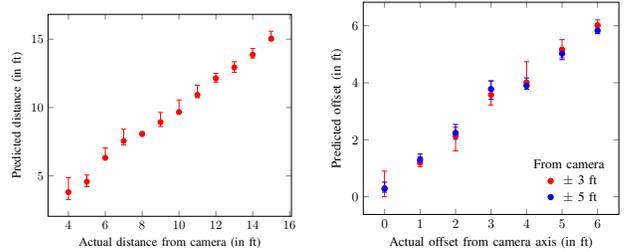
\begin{figure}[!t]
    \centering
    \subfloat[\centering Actual vs predicted distance \lFig{res-actual-vs-predicted-distance}]{{
    \resizebox{4.0cm}{!}{
    \begin{tikzpicture}
        \begin{axis} [
            xlabel=Actual distance from camera (in ft),
            ylabel=Predicted distance (in ft)
        ]
        \addplot [only marks, x=actual, y=predicted-mean, color=red] 
          plot [error bars/.cd, y dir=both, y explicit]
          table [y error plus=predicted-max, y error minus=predicted-min] {\actualVsPredictedDistance};
        \end{axis} 
    \end{tikzpicture}
    }
    }}
    \subfloat[\centering Actual vs predicted offset \lFig{res-actual-vs-predicted-offset}]{{
    \resizebox{4.0cm}{!}{
    \begin{tikzpicture}
        \begin{axis} [
            xlabel=Actual offset from camera axis (in ft),
            ylabel=Predicted offset (in ft),
            legend pos=south east,
            legend style={draw=none}
        ]
        \addlegendimage{empty legend}
        \addplot [only marks, x=actual, y=predicted-mean, color=red] 
          plot [error bars/.cd, y dir=both, y explicit]
          table [y error plus=predicted-max, y error minus=predicted-min] {\actualVsPredictedOffsetDOne};
        
        \addplot [only marks, x=actual, y=predicted-mean, color=blue] 
          plot [error bars/.cd, y dir=both, y explicit]
          table [y error plus=predicted-max, y error minus=predicted-min] {\actualVsPredictedOffsetDTwo};
        \addlegendentry{\hspace{-0.6cm}From camera}
        \addlegendentry{$\pm$ 3 ft}
        \addlegendentry{$\pm$ 5 ft}
        \end{axis}
    \end{tikzpicture}
    }
    }}
    \caption{Depth and Offset Evaluation}
    \lFig{res-depth-and-offset-evaluation}
\end{figure}
\vskip -0.1in

\subsection{Evaluation of depth \& offset estimation}

We analyzed our system with people standing at different distances and offsets from sensors and compared against ground truth. There is negative correlation between distance from sensors and difference in pixels between two cross-spectral images. As shown in \rFig{res-actual-vs-predicted-distance} and \ref{fig:res-actual-vs-predicted-offset},  observed and  model's predicted distance and offset are strongly correlated. 




\subsection{Accuracy - Edge vs Cloud}
Ability to process all the faces in the sensor's field of view reliably, depends on available bandwidth and latency between the sensor and backend system. \rTable{accuracy-of-detections} shows accuracy \cite{10.1145/1143844.1143874} of fever screening system, when the modules are deployed on a centralized cloud vs when they are deployed at the edge. Note that we measured the accuracy under same lighting and environmental conditions, with people of different age, gender, race, skin color, etc. moving either alone or together in a group. Our focus here is compare accuracy of fever screening system under same conditions, when the modules are deployed at the edge or on the cloud. We can see that the accuracy is higher at the edge compared to the cloud, because there is less network delay, which results in more frames being processed, thereby increasing the accuracy.

\begin{table}[t]
    \caption{Accuracy (Edge vs Cloud)}
    \lTable{accuracy-of-detections}
    \vskip -0.25in
    \begin{center}
    \scalebox{0.99}{
    \begin{tabular}{lccccc}
    Deployment & TP & FP & TN & FN & Accuracy \\
    \midrule
    Cloud (AWS) & 6 & 2 & 110 & 24 & 81.6\% \\
    Edge (MEC + private 5G) & 7 & 1 & 133 & 1 & 98.5\% \\
    \bottomrule
    \end{tabular}
    }
    \end{center}
\end{table}
\vskip -0.29in

\subsection{Throughput - Edge vs Cloud}
Higher the throughput i.e. number of people that are screened per minute, the better it is, since it avoids crowding, which can lead to spread of the virus. \rTable{throughput-of-detections} shows the throughput when fever screening modules are deployed on a centralized cloud vs when they are deployed at the edge, as people walk through at a normal pace of about 1 meters/s. We can see that the throughput is higher ($\sim$ 5X) when modules are deployed at the edge, since the processing for each person is much quicker due to proximity of edge resources.


\begin{table}[t]
    \caption{Throughput (Edge vs Cloud)}
    \lTable{throughput-of-detections}
    \vskip -0.25in
    \begin{center}
    \scalebox{0.99}{
    \begin{tabular}{lccccc}
    Deployment & Throughput \\
    \midrule
    Cloud (AWS) & 14 persons/min \\
    Edge (MEC + private 5G) & 66 persons/min \\
    \bottomrule
    \end{tabular}
    }
    \end{center}
\end{table}
\vskip -0.29in

\section{Related Work}
\lSec{related-work}
Methods like modified SIFT (gradient invarient SIFT- GD SIFT) \cite{Firmenichy:multispectral-image-registration}, multi-oriented Log-Gabor filters (Log-Gabor histogram descriptor - LGHD) \cite{Aguilera:lghd-feature-descriptor} and dense adaptive self-correlation (DASC) \cite{Kim:dasc-descriptor-multispectral-correspondence} 
have been proposed for cross-spectral correspondence, but these methods are not robust or accurate.
Conversion of thermal frame into visible spectrum has been explored in ThermalGAN \cite{Kniaz:thermal-gan}, 
TV-GAN \cite{Zhang:tv-gan}  and TR-GAN \cite{Kezebou:tr-gan} for cross-modal face recognition. However, these networks do not meet 
real-time constraints. 

Many stereo object localization algorithms \cite{Chang:pyramid-stereo-matching} compute pixel-wise disparity between stereo images and they generate point clouds. 
Creating cross spectral object detection models for point clouds needs lot of labelled data, which is challenging. 
In contrast, the proposed cross-spectral object depth and offset estimator generates object-level disparity maps in real-time (\rSec{depth-and-offset-estimator}).

Mobile edge computing (MEC) enables a range of applications \cite{9113305} e.g. driver-less vehicles, VR/AR, robotics and immersive media. Various private 5G enabled IoT edge applications in domain of industrial automation and healthcare are discussed in \cite{9257390}. Similar to these, we leverage MEC and 5G for an application in healthcare domain i.e. fever screening system, to identify and isolate individuals with fever. 

\section{Conclusion}

We propose an edge-based fever screening system, which leverages edge computing and 5G, and uses our novel cross-spectral generative adversarial network (CS-GAN) to achieve real-time fever screening of individuals. Our system is able to achieve 98.5\% accuracy and is able to process $\sim$ 5X more persons per minute compared to a centralized cloud deployment.

\bibliographystyle{IEEEtran}


\end{document}